\documentclass[superscriptaddress,showpacs,preprintnumbers,amsmath,amssymb,twocolumn,prl]{revtex4}

\usepackage{graphicx}% Include figure files
\usepackage{dcolumn}% Align table columns on decimal point
\usepackage{bm}% bold math
\usepackage{verbatim}% for multiline comments
\usepackage{color}
\usepackage{ulem}

\def\be{\begin{equation}}
\def\ee{\end{equation}}
\def\bea{\begin{eqnarray}}
\def\eea{\end{eqnarray}}

\renewcommand{\[}{\begin{eqnarray}}
\renewcommand{\]}{\end{eqnarray}}

\begin{document}

\title{Damping of Rabi oscillations in intensity-dependent photon echoes from exciton complexes in a CdTe/(Cd,Mg)Te single quantum well}

\author{S.~V.~Poltavtsev}
\email{sergei.poltavtcev@tu-dortmund.de}
\affiliation{Experimentelle Physik 2, Technische Universit\"at Dortmund, 44221 Dortmund, Germany}
\affiliation{Spin Optics Laboratory, St.~Petersburg State University, 198504 St.~Petersburg, Russia}
\author{M.~Reichelt}
\affiliation{Department Physik \& CeOPP, Universit\"at Paderborn, D-33098 Paderborn, Germany}
\author{I.~A.~Akimov}
\affiliation{Experimentelle Physik 2, Technische Universit\"at Dortmund, 44221 Dortmund, Germany}
\affiliation{Ioffe Physical-Technical Institute, Russian Academy of Sciences, 194021 St.~Petersburg, Russia}
\author{G.~Karczewski}
\affiliation{Institute of Physics, Polish Academy of Sciences, PL-02668 Warsaw, Poland}
\author{M.~Wiater}
\affiliation{Institute of Physics, Polish Academy of Sciences, PL-02668 Warsaw, Poland}
\author{T.~Wojtowicz}
\affiliation{Institute of Physics, Polish Academy of Sciences, PL-02668 Warsaw, Poland}
\affiliation{International Research Centre MagTop, PL-02668 Warsaw, Poland}
\author{D.~R.~Yakovlev}
\affiliation{Experimentelle Physik 2, Technische Universit\"at Dortmund, 44221 Dortmund, Germany}
\affiliation{Ioffe Physical-Technical Institute, Russian Academy of Sciences, 194021 St.~Petersburg, Russia}
\author{T.~Meier}
\affiliation{Department Physik \& CeOPP, Universit\"at Paderborn, D-33098 Paderborn, Germany}
\author{M.~Bayer}
\affiliation{Experimentelle Physik 2, Technische Universit\"at Dortmund, 44221 Dortmund, Germany}
\affiliation{Ioffe Physical-Technical Institute, Russian Academy of Sciences, 194021 St.~Petersburg, Russia}

\date{\today}

\begin{abstract}
We study Rabi oscillations detected in the coherent optical response from various exciton complexes in a 20~nm-thick CdTe/(Cd,Mg)Te quantum well using time-resolved photon echoes. In order to evaluate the role of exciton localization and inhomogeneous broadening we use selective excitation with spectrally narrow ps-pulses. We demonstrate that the transient profile of the photon echo from the localized trion (X$^-$) and the donor-bound exciton (D$^0$X) transitions strongly depends on the strength of the first pulse. It acquires a non-Gaussian shape and experiences significant advancement for pulse areas larger than $\pi$ due to non-negligible inhomogeneity-induced dephasing of the oscillators during the optical excitation. Next, we observe that an increase of the area of either the first (excitation) or the second (rephasing) pulse leads to a significant damping of the photon echo signal, which is strongest for the neutral excitons and less pronounced for the donor-bound exciton complex (D$^0$X). The measurements are analyzed using a theoretical model based on the optical Bloch equations which accounts for the inhomogeneity of optical transitions in order to reproduce the complex shape of the photon echo transients. In addition, the spreading of Rabi frequencies within the ensemble due to the spatial variation of the intensity of the focused Gaussian beams and excitation-induced dephasing are required to explain the fading and damping of Rabi oscillations. By analyzing the results of the simulation for the X$^-$ and the D$^0$X complexes we are able to establish a correlation between the degree of localization and the transition dipole moments determined as $\mu($X$^-$)=73~D and $\mu($D$^0$X)=58~D.
\end{abstract}

\maketitle

\textbf{Introduction.} Coherent control of excitonic states in semiconductor nanostructures under resonant excitation with intense optical pulses attracts a lot of attention in relation with possible applications in quantum information \cite{FoxBook}. These ideas exploit coherent rotations of the Bloch vector in the photoexcited two-level system (TLS), which depends on the area of the exciting pulse via Rabi oscillations \cite{Rabi1937, StievaterPRL2001, Zrenner2002}. Since stronger localization of excitons is in favor of longer decoherence times, most of the studies of coherent control have concentrated on quantum dots (QD) \cite{StievaterPRL2001, WangPRB2005, SuzukiPRL2016}. However, the strong localization in QDs is accompanied by large variations in QD size, shape, and composition which consequently leads to the large inhomogeneous broadening of the optical transitions when an ensemble of emitters is used. Therefore, most Rabi oscillation studies were performed on single QDs \cite{StuflerPRB2005, RamsayPRB2007, RamsayReview2010, MonnielloPRL2013}.

In semiconductor quantum well (QW) structures the inhomogeneous broadening of the optical transitions is significantly smaller as compared to QD systems, i.e. it is possible to selectively address different exciton complexes, such as free and localized excitons, localized charged excitons (trions, X$^-$), and donor-bound excitons (D$^0$X). Therefore, QW structures can be considered as a model system for the investigation of Rabi oscillations and their damping for optical excitations with different degree of localization and inhomogeneity. In spite of this fact, so far only few studies on Rabi oscillations have been performed in QW structures \cite{GibbsPRL1999,LangbeinPRL2005}. This is mainly due to the presence of many-body interactions, which can, however, be suppressed by using spectrally-narrow optical pulses \cite{NollPRL1990, LangerPRL2012, LangerNature}. Another important issue is related with the detection of Rabi oscillations. One of the approaches is based on reading out of the excited state population \cite{StievaterPRL2001, Zrenner2002, GibbsPRL1999},  while another exploits direct measurements of the coherent optical response \cite{LangbeinPRL2005, SuzukiPRL2016}. Due to the inhomogeneous broadening of optical transitions in the TLS ensemble, the coherent optical response in a four-wave-mixing (FWM) experiment is represented by photon echoes \cite{BermanMalinovskyBook}. In contrast to single pulse excitation, in FWM the amplitude and the transient profile of the detected signal depend sensitively on the optical field amplitude $A_i$ (pulse area $\Theta_i$) of both exciting ($i = 1$) as well as rephasing ($i = 2$) pulses which, as is demonstrated below, provides valuable information on the relevant physical effects.

In this work, we study the coherent optical response for intense resonant excitation of various exciton complexes with different degrees of localization and inhomogeneity in a single CdTe/(Cd,Mg)Te QW structure. In our low temperature measurements we observe photon echoes (PE) from the ensembles of individually addressed states of localized excitons, trions, and donor-bound excitons. By scanning the areas of the incident pulses and measuring the temporal profiles of photon echo for each exciton complex we observe Rabi oscillations in the form of two-dimensional images similar to those observed recently in an ensemble of InAs/(In,Ga)As quantum dots \cite{PoltavtsevPRB2016}. The images vary for the different complexes because they are very sensitive to the degree of inhomogeneous broadening of the optical transition. Besides, the photon echoes are strongly damped for large pulse areas. In order to explain this damping, we develop a theoretical model, which takes into account the two most important damping mechanisms: (i) Excitation-induced dephasing (EID), which results in the accelerated decay of the coherence with increasing excitation intensity, and (ii) a fading of the Rabi oscillations due to the spatial distribution of the optical excitation. Our results demonstrate that the ensemble of D$^0$X complexes in CdTe QW structures with low donor concentrations ($n \leq 10^{10}$ cm$^{-2}$) is a fairly good two-level system for the generation of intense photon echoes with the $\pi/2-\pi$ pulse area sequence. However, larger excitation densities lead to unavoidable many-body effects and heating of the electron system which destroy the coherence and diminish the photon echo signals.

The paper is organized as follows. First, we describe the experimental technique that is used to measure Rabi oscillations and give details of the studied sample. Then, the experimental results are presented followed by the theoretical model used to analyze the measurements. Finally, a discussion and conclusions are provided.

\begin{figure}[t]
	\vspace{5mm}
	\includegraphics[width=\linewidth]{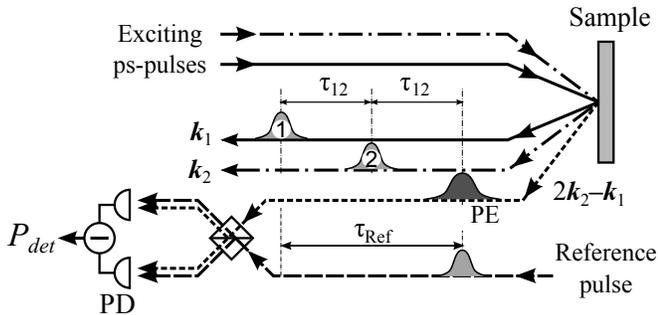}
	\caption{Scheme of PE detection in reflection geometry. PD denotes the photodetector.}
	\label{setup}
\end{figure}

\textbf{Experimental method and sample.} In order to investigate photon echoes we used the time-resolved degenerate four-wave-mixing (FWM) technique with optical heterodyning. The optical excitation of the sample was performed using picosecond pulses emitted by a tunable Ti:Sapphire laser with a repetition rate of 75.75~MHz. The sample immersed in liquid helium and cooled down to the temperature of 1.8~K was excited by a sequence of two laser pulses separated by a variable delay $\tau_{12}$ hitting the sample under the incidence angles of 3$^\circ$ and 4$^\circ$ ($\bm{k}_1$ and $\bm{k}_2$), respectively, and focused on the sample to a spot with diameter of about 300~$\mu$m. The FWM signal was collected in reflection geometry in the $2\bm{k}_2-\bm{k}_1$ direction, as schematically shown in Fig.~\ref{setup}. This signal was mixed with the optical field of a reference pulse delayed with respect to the first pulse by $\tau_\text{Ref}$ at the balanced detector using a non-polarizing beamsplitter. All laser pulses including the both exciting pulses and the reference pulse were linearly co-polarized. For heterodyne FWM measurements, the optical frequencies of the first exciting beam and the reference beam were shifted using acousto-optical modulators by $-41$~MHz and $+40$~MHz, respectively. The modulus of the cross-correlation of the FWM amplitude $P_{FWM}$ with the Gaussian reference pulse field detected at the frequency of 1~MHz is described by

\be
P_{det}(\tau_{Ref})\sim\bigg|\int\limits_{-\infty}^{+\infty}P_{FWM}^*(\tau_{Ref}-t')\cdot \exp(-t'/2\tau_P^2)dt'\bigg|,
\label{P_WFM}
\ee

\noindent Here, $2\sqrt{\ln2}\tau_P=2.2$~ps is the laser pulse duration in the intensity scale.

\begin{figure}[h]
	\vspace{5mm}
	\includegraphics[width=\linewidth]{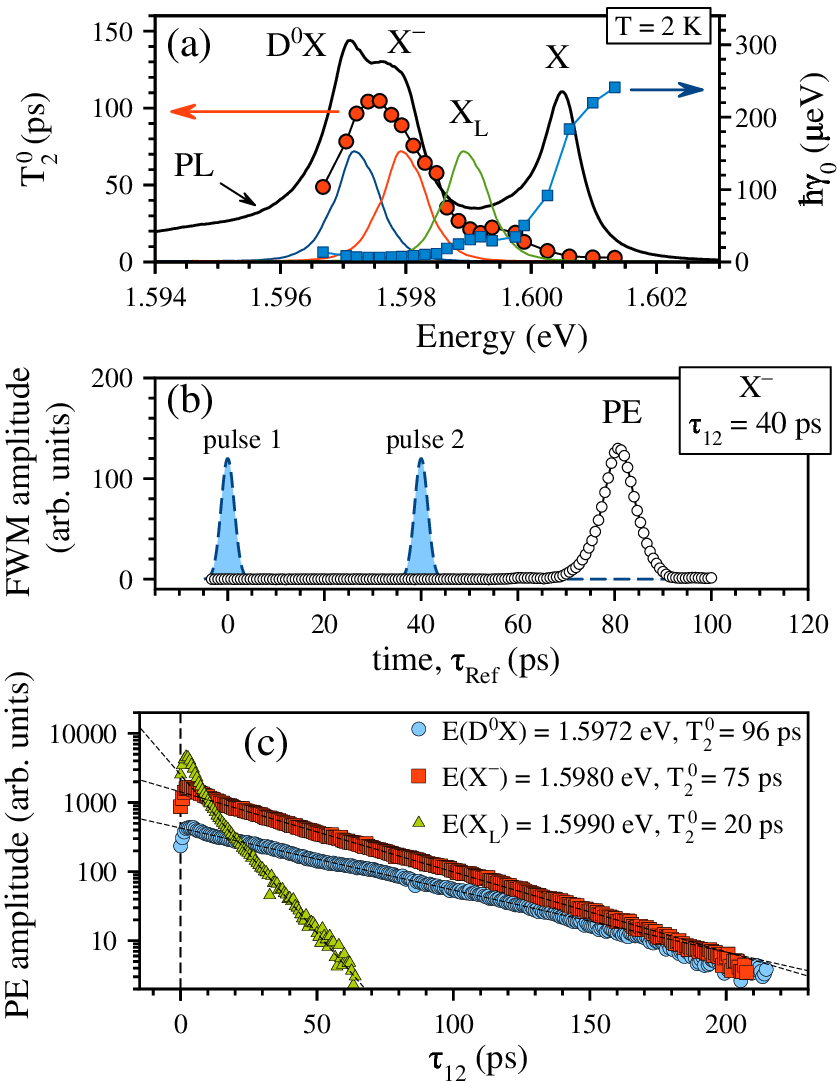}
	\caption{(Color)  (a) PL spectrum of the studied sample (black line) with indicated spectral features: donor-bound exciton at E(D$^0$X)=1.5972~eV, trion at E(X$^-$)=1.5980~eV, and neutral exciton at E(X)=1.6005~eV. Spectra of laser pulses tuned to D$^0$X, X$^-$, and the localized exciton state (X$_\text{L}$) are shown by the blue, red, and green line, respectively. Red dots are the coherence times $T_2^0$ and blue dots are the homogeneous linewidths $\hbar\gamma_0$ obtained from scans of the PE decays. (b) Transient of FWM signal measured at the trion at $\tau_{12}=40$~ps demonstrating a PE located at $\tau_{Ref}=2\tau_{12}=80$~ps (circles). The exciting pulses are schematically shown by the dashed line. (c) Decays of the PE amplitude measured at three indicated energies with extracted coherence times $T_2^0$.}
	\label{Sample}
\end{figure}

To study the coherent optical dynamics of the localized systems we chose a structure with a high quality 20-nm thick single CdTe QW grown by molecular beam epitaxy. The structure was grown on (100)-oriented GaAs substrate followed by a 4.5~$\mu$m-thick Cd$_{0.76}$Mg$_{0.24}$Te buffer and 5 short-period superlattices separated by 100~nm CdMgTe spacers. This is followed by the QW sandwiched between the 100~nm CdMgTe barriers. The structure was not intentionally doped with donors. The resident electron density due to an unavoidable background of impurities is estimated to be $n_e \leq 10^{10}$ cm$^{-2}$. Recent studies performed on this sample have demonstrated the presence of two very closely located optical transitions corresponding to the negatively charged trion (X$^-$) and the neutral donor-bound exciton (D$^0$X) at 1.5980~eV and 1.5972~eV, respectively \cite{AkimovArxive2017}, which are clearly seen in the photoluminescence (PL) spectrum measured at a temperature of 2~K, see Fig.~\ref{Sample}(a). Additionally, it displays the neutral exciton line (X) located at 1.6005~eV. In order to increase the density of resident electrons, which are required for the excitation of the trion, an additional weak above-barrier-illumination from a spectrally broad white light source was applied.

\begin{figure*}[t]
	\vspace{5mm}
	\includegraphics[width=\linewidth]{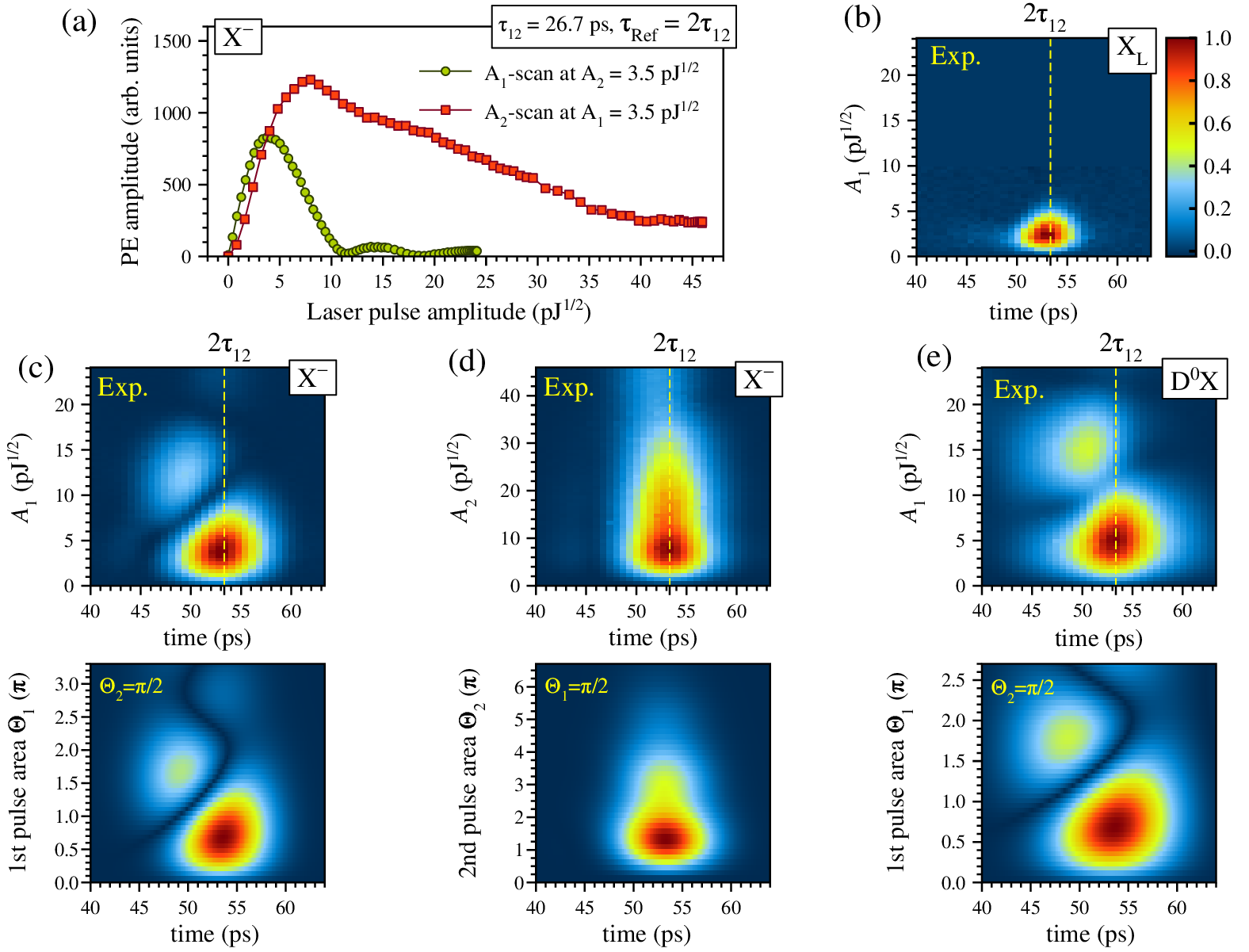}
	\caption{(Color) Rabi oscillations measured on the various localized exciton complexes in the studied CdTe/(Cd,Mg)Te QW. (a) Dependence of PE amplitude measured on the trion (E=1.5980~eV) detected at $\tau_{Ref}=2\tau_{12}=53.3$~ps on both pulse amplitudes. Measurements of PE transients as a function of pulse amplitude: localized exciton at E(X$_\text{L}$)=1.5990~eV, $A_1$ is varied at $A_2=3.5$~pJ$^{1/2}$ (b); localized trion at E(X$^-$)=1.5980~eV, $A_1$ is varied at $A_2=3.5$~pJ$^{1/2}$ (c), $A_2$ is varied at $A_1=3.5$~pJ$^{1/2}$ (d); donor-bound exciton at E(D$^0$X)=1.5972~eV, $A_1$ is varied at $A_2=3.4$~pJ$^{1/2}$ (e). Upper and lower panels of (c)-(e) are the experimental data (upper row) and the numerical simulations (lower row), respectively. The scales in panels (b)-(e) are normalized to unity.}
	\label{Rabi_Lambda}
\end{figure*}

\textbf{Experimental results.} Figure~\ref{Sample}(b) displays a typical FWM transient measured at $\tau_{12}=40$~ps, in which a single PE peak at $2\tau_{12}=80$~ps is seen. To obtain the optical coherence time $T_2^0$, the delay $\tau_{12}$ was varied and the PE amplitude was detected at $\tau_{Ref}=2\tau_{12}$ in the regime of weak optical excitation. This procedure was performed for a number of spectral positions, and Fig.~\ref{Sample}(c) displays three PE decays measured on the donor-bound exciton at E(D$^0$X)=1.5972~eV, the trion at E(X$^-$)=1.5980~eV, and the localized exciton state at E(X$_\text{L}$)=1.5990~eV. These decays are well fitted by the exponential decay function $\sim\exp(-\tau_{12}/2T_2^0)$ which provides the following values of the coherence time: $T_2^0($D$^0$X)=96~ps, $T_2^0($X$^-$)=75~ps, $T_2^0$(X$_\text{L}$)=20~ps, and corresponding homogeneous linewidth of $\hbar\gamma_0=\hbar/T_2^0$, respectively: $\hbar\gamma_0($D$^0$X)=6.9~$\mu$eV,  $\hbar\gamma_0($X$^-$)=8.8~$\mu$eV, and $\hbar\gamma_0($X$_\text{L}$)=33~$\mu$eV. The spectral dependencies of $T_2^0$ and $\hbar\gamma_0$, shown in Fig.~\ref{Sample}(a), are measured in a similar way. From these data we see that the donor-bound exciton and trion have long coherence times up to 100~ps, while the free excitons located at energies above 1.6005~eV reveal a very short $T_2^0$ of below 10~ps. The optical transitions in the region of 1.599-1.600~eV have moderate coherence times on the order of 20~ps, which we attribute to localized exciton states. It is worth noting that the low-energy side of the donor-bound exciton below 1.5975~eV shows some shortening of the coherence times down to 50~ps, while the exciton and the trion show an increase of $T_2^0$ with decreasing excitation energy. The degree of the exciton and the trion localization on the potential and composition fluctuations increases with the decrease of the energy. Thus, longer $T_2^0$ times are expected for lower energy excitons and trions \cite{NollPRL1990}. For D$^0$X, the transition energy is known to be affected by other factors, in particular, the interaction between neighboring donors \cite{EfrosBook}. Thus, the spectral dependence of $T_2^0$ for D$^0$X can be more complex and is clearly different from that of the localized exciton and trion.

In order to observe Rabi oscillations, PE transients similar to that shown in Fig.~\ref{Sample}(b) were recorded as a function of the amplitude of one of the two exciting pulses, $A_i$ ($i=1,2$), which we define as the square root of the energy per pulse, while that of the other pulse was kept constant. Thus, we vary the exciting pulse areas $\Theta_i$ ($i=1,2$) and keep the pulse duration $\tau_P$ fixed. Figure~\ref{Rabi_Lambda} summarizes the experimental results obtained at the three spectral positions indicated above.

The excitation of the localized exciton at E(X$_\text{L}$)=1.5990~eV results in a PE positioned at $2\tau_{12}\approx53$~ps, as shown in Fig.~\ref{Rabi_Lambda}(b). Here, the first pulse amplitude $A_1$ was scanned using a constant $A_2=3.5$~pJ$^{1/2}$. The PE amplitude becomes damped at $A_1\approx7$~pJ$^{1/2}$ and no oscillatory behavior can be observed. This behavior is to be expected since the nonlinear response of free and weakly localized excitons is very strongly affected by many-body interactions that lead to rapid dephasing and thus prevent the appearance of oscillatory intensity-dependent echo signals \cite{WangPRL1993, JahnkePRL1996, Kochbook}.

The same type of measurement performed on the trion at E(X$^-$)=1.5980~eV and shown on the upper panel of Fig.~\ref{Rabi_Lambda}(c) displays a drastically different result. When the first pulse amplitude $A_1$ is varied at $A_2=3.5$~pJ$^{1/2}$, the PE transients reveal a complex two-dimensional picture with the split PE profile similar to that observed recently in (In,Ga)As quantum dots \cite{PoltavtsevPRB2016}. The first maximum of the Rabi oscillations for $A_1=3.6$~pJ$^{1/2}$ is centered at $2\tau_{12}\approx53$~ps and well described by a single pulse of Gaussian shape, which is expected for a Hahn echo \cite{AllenEberly}. However, the second maximum at $A_1=13$~pJ$^{1/2}$ is significantly advanced. The tail of the third maximum of Rabi oscillations at $A_1>20$~pJ$^{1/2}$ appears to be non-shifted. When the amplitude of the second pulse $A_2$ is varied at $A_1=3.5$~pJ$^{1/2}$, the PE maximum has no shift, as can be seen from the upper panel of Fig.~\ref{Rabi_Lambda}(d). Here, however, only the first maximum of Rabi oscillations at $A_2=7.5$~pJ$^{1/2}$ is pronounced, while the second one around $A_2=20$~pJ$^{1/2}$ is very weak. This is better visible in Fig.~\ref{Rabi_Lambda}(a), where the cross-sections of the two-dimensional plots of Rabi oscillations according to the variations of both pulse amplitudes are plotted.

Figure~\ref{Rabi_Lambda}(e) displays the measurement of Rabi oscillations performed on the donor-bound exciton at E(D$^0$X)=1.5972~eV. Here, when $A_1$ is varied at $A_2=3.4$~pJ$^{1/2}$, the PE transients appear somewhat broader in time as compared to those measured on the trion. This means that the optically-excited inhomogeneous ensemble of D$^0$X is narrower than that for the trion. Also, the transition dipole moment of D$^0$X is smaller than that of X$^-$, since the Rabi frequency is smaller.

\textbf{Theoretical model and discussion.} The experimental measurements clearly manifest a rich intensity-dependent coherent behavior of the donor-bound exciton and the trion, which can be analyzed by modeling them as inhomogeneous ensembles of two-level systems interacting with the optical field.
This approach is, however, unlikely to be applicable to the localized excitons X whose dynamics is more strongly influenced by complex many-body correlations \cite{WangPRL1993, JahnkePRL1996, Kochbook} that may not be described properly by the modeling used here.
In the following we develop an adapted theoretical model that is able to describe the observed intensity-dependence of PEs measured on D$^0$X and X$^-$ and study the mechanisms responsible for the observed damping of the Rabi oscillations.

%------

To simulate the photon echo transients, we numerically solve sets of the optical Bloch equations~\cite{AllenEberly,TMKbook}, i.e.,
the coupled equations of motion for the microscopic polarization $p$ and the occupation of the excited state $n$,
considering excitation parameters corresponding to the experimental situations. Since an adequate explanation of the experimental results requires the incorporation of a number of effects, which are described in more detail below, the numerically-solved system of equations contains two indices $f$ and $s$:
\[
\frac{\partial}{\partial t} \,p_{f,s}(t) &=& \big( -i \, \omega_{f} - \gamma (t) \big) \, p_{f,s}(t) \nonumber \\
&& + \, i \frac{\mu}{\hbar} E_{s}(t) \big(1-2n_{f,s}(t)\big)
\label{Blocheqs1} \, , \\
\frac{\partial}{\partial t} \,n_{f,s}(t) &=& -2 \frac{\mu}{\hbar} \,
\mathrm{Im} \big[ p_{f,s}(t)^* E_{s}(t)\big] \, .
\label{Blocheqs2}
\]
Here, $\omega_f$ is the optical transition frequency of the TLS, $\mu$ is the dipole matrix element which is taken to be constant for the ensemble, $E_s$ the total electric field including both laser pulses, $E(t)=E_1(t)+E_2(t)$, and $\gamma(t)$ is the excitation and thus time-dependent dephasing rate. Even a qualitative explanation of the experimental findings requires to incorporate three effects into the theoretical analysis: (i) An inhomogeneous broadening of the resonance, i.e., a Gaussian distribution of the optical frequency within the TLS ensemble which is described by the $f$; (ii) the spatial profile of the incident laser pulses which is described as a Gaussian leading to the additional index $s$; and (iii) excitation-induced dephasing, which is caused by the amount of optical excitation and, therefore, leads to the time-dependent dephasing rate $\gamma(t)$.

By solving the system of equations~(\ref{Blocheqs1}) and (\ref{Blocheqs2}) we compute the total FWM polarization describing photon echoes as
\[
P_{FWM}(t) =\sum_{f,s}
\mu\,\alpha_f\,\beta_s\,p_{f,s}(t)\,,
\label{Pmacr}
\]
\noindent where $\alpha_f$ and $\beta_s$ are weight coefficients described further below. To compare with the experimentally-detected signal we convolute Eq.~(\ref{Pmacr}) with the Gaussian shaped reference pulse, as described by Eq.~(\ref{P_WFM}).

\begin{figure*}[t]
	\vspace{5mm}
	\includegraphics[width=\linewidth]{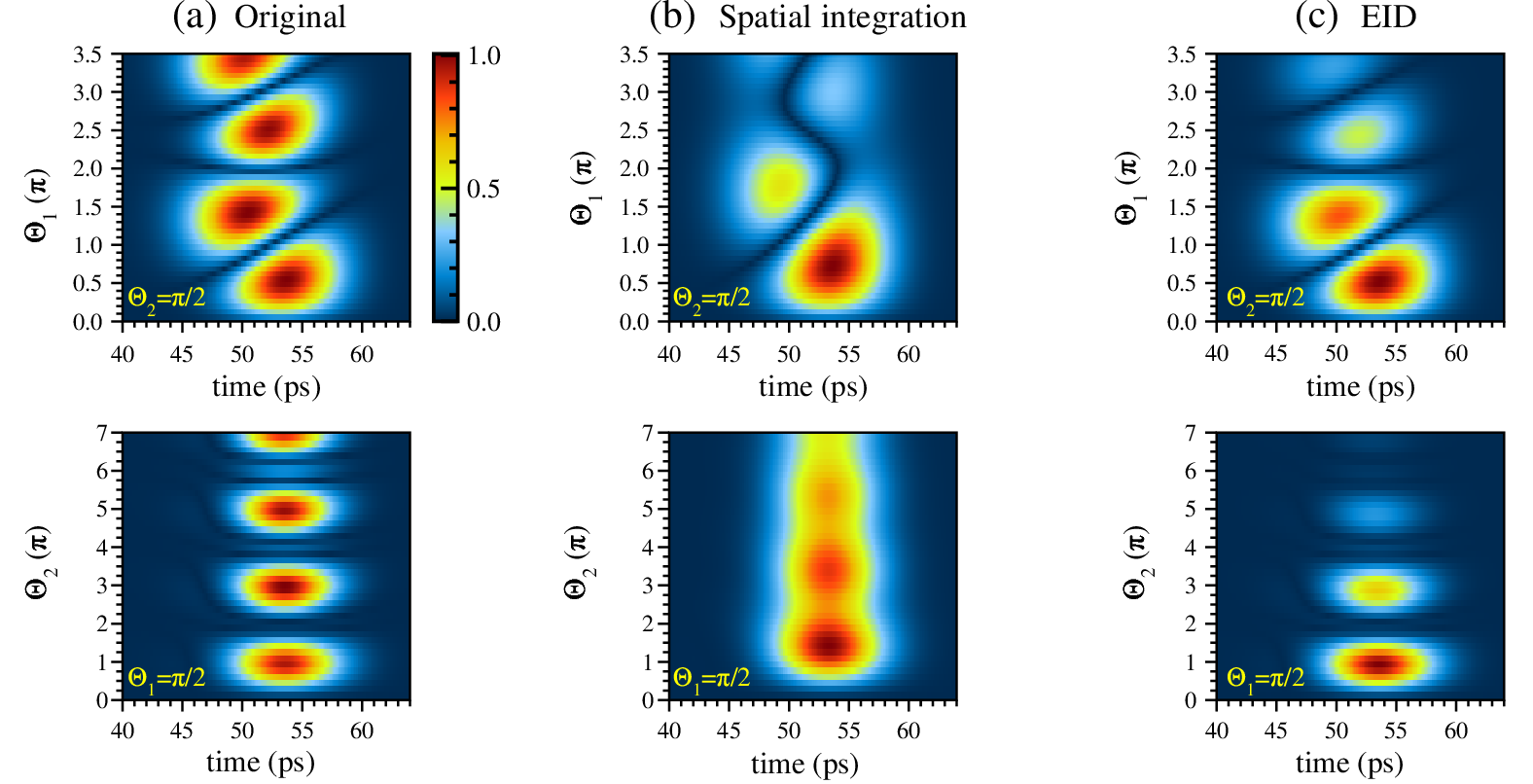}
	\caption{(Color) Various mechanisms responsible for the damping of the simulated Rabi oscillations: (a) Rabi oscillations without damping mechanisms. (b) Only the spatial integration over the excitation beam is activated and $\gamma = 1/T_2^0$. (c) Only EID is activated. Upper panels correspond to the first pulse area scan at $\Theta_2=\pi/2$; lower panels correspond to the second pulse area scan at $\Theta_1=\pi/2$. All intensity scales are normalized to unity. The width of the spectral ensemble used in the calculations is $2\sqrt{2\ln2}\sigma=1.0$~meV and the pulse delay is $\tau_{12}=26.7$~ps.}
	\label{Rabi_mechanisms}
\end{figure*}

When setting up and solving Eqs.~(\ref{Blocheqs1})-(\ref{Blocheqs2}), we take a number of experimental requirements into account: (i) In order to obtain a photon echo, an inhomogeneous distribution of the optical transition energies is necessary. We define the spectral density of oscillators by a Gaussian distribution $\alpha_f\propto \exp[-(\omega_f-\omega_0)^2/2\sigma^2]$ centered at $\omega_0$, which is coincident with the optical field frequency. Here, we adjust the width $\sigma$, such that the temporal profile of the echo calculated at both exciting pulse areas $\Theta_i=\int \frac{\mu}{\hbar}E_i(t)dt=\pi/2$, ($i=1, 2$) \cite{AllenEberly}, and the temporal profile of the PE measured at $A_1\approx A_2\approx3.5$~pJ$^{1/2}$ have equal temporal widths. 

(ii) The spread of Rabi frequencies within the ensemble. A statistical distribution of the dipole moments in the ensemble can result in a variation of the pulse area. However, in contrast to strongly inhomogeneous QD ensembles, this effect is unlikely to take place in a QW structure \cite{LangbeinPRB2002,SalewskiPRB2017}. Here, we, however, have to consider the spatial variation of the pulse within the excitation spot which is modeled by a Gaussian profile $E_{s}\sim\exp(-r^2/2\sigma_R^2)$ in real space ($r$ is the distance from the laser spot center), with $\sigma_R$ characterizing the size of the focused laser beam. This spatial excitation profile defines also a weight $\beta_s\sim r$, which corresponds to the amount of oscillators, located in the sample area with the radius $r$ and excited by the field amplitude $E_s$. In other words, due to the nonlinear characteristics of the Bloch equations~(\ref{Blocheqs1})-(\ref{Blocheqs2}), various Rabi frequencies are superimposed and enter the total signal with according weight $\beta_s$. This spatial integration has a profound effect on the $\Theta_1$ and $\Theta_2$ dependencies and leads to an additional decrease of the Rabi oscillation amplitude for the higher pulse areas. This is visualized in Fig.~\ref{Rabi_mechanisms}(a) and (b), where numerically simulated Rabi oscillations are shown for the cases, when all dephasing mechanisms are deactivated and only the spatial integration is considered, respectively. Additionally to the damping, the effective frequency of the Rabi oscillations is also affected. It should be noted that the result of the calculations does not depend on the value of the beam size $\sigma_R$, when the spatial integration is performed for a large enough area ($\max(r)\gtrsim3\sigma_R$). 

(iii) The dephasing rate $\gamma(t)=\gamma(E(t))$, which depends on the state of the optical excitation of the system and thus on the exciting electric fields, is a crucial factor in modeling the different experimental situations. It is known that Rabi oscillations can be damped depending on the excitation strengths of the driving fields and quite a few different mechanisms have been identified. The damping might be due to phonons, population leakage into the delocalized states, Auger capture or the transfer of the excitation to other states \cite{LangbeinPRL2005, GreenlandNature2010, KruegelApb2005, Zhou2005}. For the trion and donor-bound exciton heating of the electronic ensemble in the ground state is also one of the  relevant mechanisms \cite{ZhukovPRB2010}. Nevertheless, it has been shown that the concrete mechanism is less important than the non-Markovian behavior of a reservoir~\cite{Mogilevtsev2009, Mogilevtsev2008}. Here, we assume that for our conditions the intensity-dependence of the dephasing rate is governed by excitation-induced dephasing (EID)\cite{WangPRL1993, JahnkePRL1996}.
In our simulations we describe this dependence by
\[
\gamma (t) = \frac{1}{T_2^0} + a \int_{-\infty}^t \,dt' E^2(t') \,.
\label{EID_eq}
\]
The main idea is that the laser field off-resonantly excites populations that via scattering effects lead to a damping of the polarization. The effect of EID can be seen in Fig.~\ref{Rabi_mechanisms}(c), where Rabi oscillations are simulated with activated EID without integrating over the spatial coordinate.

\begin{figure}[t]
	\vspace{5mm}
	\includegraphics[width=\linewidth]{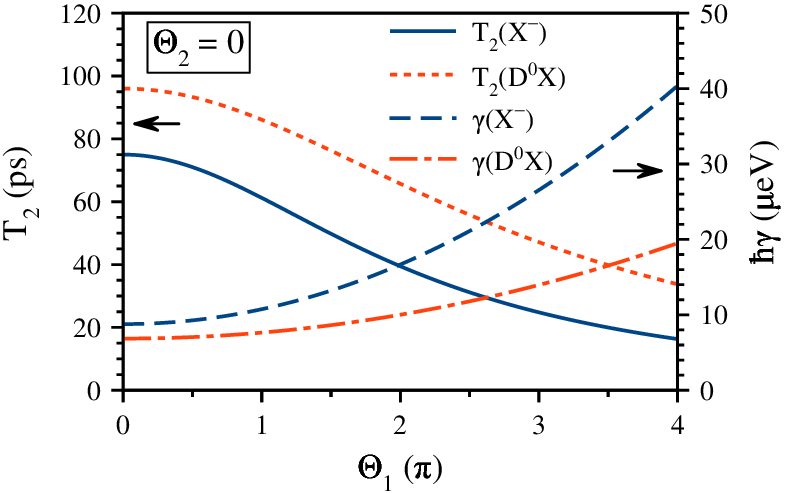}
	\caption{(Color) Calculated excitation-induced dephasing as a function of the first pulse area $\Theta_1$ at $\Theta_2=0$: solid and dotted lines are the dephasing times $T_2=1/\gamma$ of X$^-$ and D$^0$X, respectively; dashed and dash-dotted lines are the dephasing rates $\gamma$ of X$^-$ and D$^0$X, respectively.}
	\label{EID_result}
\end{figure}

By adjusting the model parameters for the effects (i)-(iii), using $T_2^0$ obtained from the PE decay measurements with weak pulses as well as the proper laser pulse characteristics, one can model and understand the measured echo signals shown in Fig.~\ref{Rabi_Lambda}(c)-(e). Results of the Rabi oscillation simulations are shown in this figure in the lower panels under the according experimental plots. In these simulations, the adjusted spectral widths of the X$^-$ and D$^0$X ensembles are $2\sqrt{2\ln2}\sigma=1.0\pm0.1$~meV and $0.5\pm0.1$~meV, respectively. This ratio of linewidths seems reasonable, since the potential for donor-bound exciton is less affected by the fluctuations of composition and QW width. The adjusted EID parameter $a$ is different in the two simulations, so that $a$(X$^-$)/$a$(D$^0$X)=2.5 giving significantly different EID for the two considered exciton complexes. It can be seen from Fig.~\ref{EID_result}, where the calculated dephasing rate $\gamma$ and the dephasing time $T_2=1/\gamma$ are plotted as functions of the first pulse area $\Theta_1$ at $\Theta_2=0$. As a result, the Rabi oscillations for the case of donor-bound excitons are more robust. This qualitative conclusion correlates with the assumption that D$^0$X is stronger localized than the trion and, hence, many-body interactions leading to EID are weaker for this transition.

%\cck{It should be noted that our consideration of the EID mechanism using Eq.~\ref{EID_eq} is simplified and provides a rather qualitative description of the data in Fig.~\ref{Rabi_Lambda}(a) and (b). A much better agreement of the simulation with the measurement can be obtained using a more phenomenological approach, in which the intensity-dependent part of the EID rate $\gamma$ differently depends on each beam intensity: $\gamma_\text{EID}\sim b\,E_1^2+c\,E_2^2$, where $b/c\approx3$ giving a less manifested EID for the second pulse intensity scan. Although this approach is hard to justify, it can provide more quantitative data description.}

Finally, from the Rabi oscillation simulations we can obtain the transition dipole moments for both resonances. In order to calculate them we use the expression for the Rabi frequency, $\Omega_R=\mu E / \hbar$, and take into account the laser spot size, the pulse duration, and the scale of the Rabi oscillations in units of the pulse amplitude known from the experimental data. As a result, we evaluate $\mu$(X$^-)\approx 73$~D and $\mu$(D$^0$X$)\approx 58$~D. This gives a ratio of $\mu$(X$^-) / \mu$(D$^0$X) = 1.26 which can also be deduced from the period of the Rabi oscillations in Fig.~\ref{Rabi_Lambda}(c) and (e). We can compare this value with another estimation based on first-principles calculations, from which we know that $\mu^2\sim \tau_0$, where $\tau_0$ is the radiative lifetime of the optical excitation \cite{IvchenkoBook}. Values of the optical lifetime $T_1$ for X$^-$ and D$^0$X were measured earlier in the same sample at low temperature: $T_1$(D$^0$X)=66~ps, $T_1$(X$^-$)=55~ps \cite{AkimovArxive2017}. Comparing these values with $T_2^0$(D$^0$X)=96~ps and $T_2^0$(X$^-$)=75~ps measured here we have the ratio $2T_1\approx T_2^0$ roughly fulfilled for both transitions for weak optical excitation. This means that the energy decay of both the trion and the donor-bound exciton is mainly radiative, i.e. $\tau_0\approx T_1$. Thus, we can estimate $\mu$(X$^-) / \mu$(D$^0$X)$\approx\sqrt{T_1(\text{D}^0\text{X}) / T_1(\text{X}^-)}$ = 1.10, which is in qualitative agreement with the previous estimation. We can therefore conclude that the transition dipole moment of the trion and the donor-bound exciton correlates with the degree of localization of these particles.

%As a result, our findings provide a spectroscopic method to study optical coherent evolution of various localized optically excited states in the semiconductor nanostructures. Our technique of the picosecond photon echo-based detection of the Rabi oscillations provides a measurement, which is highly sensitive to the specific properties of the spectrally addressed transition. It gives a rich picture of temporal coherent evolution of this transition, which can be understood using the proposed model. With this technique, such properties as inhomogeneous spectral width, dipole transition moment and excitation-induced dephasing, additionally to the decoherence time $T_2^0$ provided by PE decay measurement, can be studied on optically accessible exciton complexes.

\textbf{Conclusion.} We have observed and analyzed optical Rabi oscillations by measuring photon echoes from the trion and the donor-bound excitons in a CdTe/(Cd,Mg)Te QW structure. The photon echoes detected from the localized exciton states exhibit strong excitation-induced dephasing, which rapidly quenches the Rabi oscillations. Our experimental findings together with the proposed model provide a spectroscopic method by which the coherent evolution of various optical excitations can be studied in detail. By comparing the results of numerical simulations for the D$^0$X and the X$^-$ complexes we establish a correlation between the degree of localization and the transition dipole moment of the exciton complexes: the stronger localized donor-bound exciton has a smaller $\mu$ than the trion (58~D and 73~D, correspondingly). It follows that the influence of EID on the coherent response of D$^0$X is significantly weaker as compared to the localized trion ($\hbar\gamma$=9~$\mu$eV and 14~$\mu$eV, respectively, at $\Theta_1 = 3\pi/2$, $\Theta_2 = \pi/2$). The inhomogeneous broadening extracted from the simulations amounts to 0.5~meV and 1.0~meV for  D$^0$X and X$^-$, respectively. The experiment and the simulations demonstrate that the Rabi oscillations are strongly smoothed due to the spatially inhomogeneous optical excitation spots. This averaging can be significantly reduced or overcome by using a spatial mask to define a more homogeneously excited sample area, which we consider as a prospect for future studies.

{\it Acknowledgments.} We acknowledge financial support of the Deutsche Forschungsgemeinschaft (DFG) through the Collaborative Research Centre TRR 142 (project A02) and the International Collaborative Research Centre 160, the latter of which is also supported by the Russian Foundation of Basic Research (project N 15-52-12016 NNIO$\_$a). M.B. acknowledges the partial financial support from the Russian Ministry of Science and Education (contract no. 14.Z50.31.0021). The research in Poland was partially supported by the National Science Centre (Poland) through Grants No. DEC-2012/06/A/ST3/00247 and No. DEC-2014/14/M/ST3/00484, as well as by the Foundation for Polish Science through the IRA Programme co-financed by EU within SG OP.

%\bibliography{CdTeRabiEID}

\end{document}